%% file: main.tex
\documentclass[sigconf,nonacm]{acmart}

\newcommand{\leakfuzzer}{LeakFuzzer}

\usepackage[shortlabels]{enumitem}
\usepackage{multirow, makecell, booktabs}
\usepackage{fancyvrb}

\usepackage{xcolor}	
\usepackage{soul}

\AtBeginDocument{%
	}

\setcopyright{none}
\copyrightyear{2023}
\acmYear{2023}
\acmDOI{XXXXXXX.XXXXXXX}

\begin{document}
	
	\title{Hyperfuzzing: black-box security hypertesting with a grey-box fuzzer}
	
	%% \authornote{Both authors contributed equally to this research.}

	\author{Daniel Blackwell}
	\email{d.blackwell@cs.ucl.ac.uk}
	\affiliation{%
		\institution{University College London}
		\streetaddress{}
		\city{London}
		\country{UK}
	}

	\author{Ingolf Becker}
	\email{i.becker@cs.ucl.ac.uk}
	\affiliation{%
		\institution{University College London}
		\streetaddress{}
		\city{London}
		\country{UK}
	}

	\author{David Clark}
	\email{david.clark@ucl.ac.uk}
	\affiliation{%
		\institution{University College London}
		\streetaddress{}
		\city{London}
		\country{UK}
	}

	\renewcommand{\shortauthors}{Blackwell et al.}
	
	\input{abstract.tex}
	

	
	
	\maketitle

	\input{Introduction_v2.tex}

	\input{background.tex}
	\input{technique.tex}

	\input{evaluation.tex}

	\input{conclusion.tex}

	\bibliographystyle{ACM-Reference-Format}
	\bibliography{references}

\end{document}

%% file: abstract.tex
\begin{abstract}
Information leakage is a class of error that can lead to severe consequences.
However unlike other errors, it is rarely explicitly considered during the software testing process.
 \leakfuzzer~ advances the state of the art by using a noninterference security property together with a security flow policy as an oracle. As the tool extends the state of the art fuzzer, AFL++, \leakfuzzer~  inherits the advantages of AFL++ such as scalability, automated input generation, high coverage and low developer intervention.

The tool can detect the same set of errors that a normal fuzzer can detect, with the addition of being able to detect violations of secure information flow policies. 

We evaluated \leakfuzzer{} on a diverse set of 10 C and C++ benchmarks containing known information leaks, ranging in size from just 80 to over 900k lines of code.
Seven of these are taken from real-world CVEs including Heartbleed and a recent error in PostgreSQL.
Given 20 24-hour runs, LeakFuzzer can find 100\% of the leaks in the SUTs whereas existing techniques using such as the CBMC model checker and AFL++ augmented with different sanitizers can only find 40\% at best.
\end{abstract}

%% file: Introduction_v2.tex
\section{Introduction}

In 2015 Johannes Kinder made the case for \emph{automated} hypertesting \cite{kinder2015hypertesting}. His motivating example was Heartbleed, the famous violation in OpenSSL of the now standard program flow security hyperproperty \cite{ClarksonS2008} called \emph{non-interference}  \cite{gogmes}. Until our creation of Leakfuzzer, Kinder's implicit challenge had not been met. LeakFuzzer is the first tool that automatically generates a set of hypertests and uses them to check that a C/C++ program together with its security policy do not violate the non-interference property. Since LeakFuzzer, as its name implies, is a modification of the greybox mutational feedback fuzzer AFL++ \cite{fioraldi2020afl++}, it not only automatically generates new test inputs but it inherits the fuzzer's exploratory power with respect to branch coverage of the target program. 

The key theoretical underpinning for LeakFuzzer is the non-interference property. In Goguen and Meseguer's 1982 paper the property was stated in a general way:\\
\emph{One group of users, using a certain set of commands, is \emph{noninterfering} with another group of users if what the first group does with those commands has no effect on what the second group of users can see} \cite{gogmes}. \\
There are many ways this idea can be instantiated and for \leakfuzzer~ we use a version of this property that frequently occurs in the secure flow literature \cite{SabMy03}. 
We only observe program inputs and outputs to determine whether leakage has occurred (hence our characterisation of the security property as \emph{``black box''}). 
Further, the property only considers terminating inputs and assumes that each input has at most both a high security part and a low security part, similarly for outputs. 
For example, the interface through which Heartbleed is triggered has a low security input and a low security output, but the memory accessed by the process during execution must be considered high security as keys or other secrets may be stored there. 
This (low security inputs and outputs with memory protected as a high input) is a common security policy in practice for multi-user programs / systems \cite{heusser2010quantifying}. 

\begin{figure}
	\centering
	\includegraphics[width=\columnwidth]{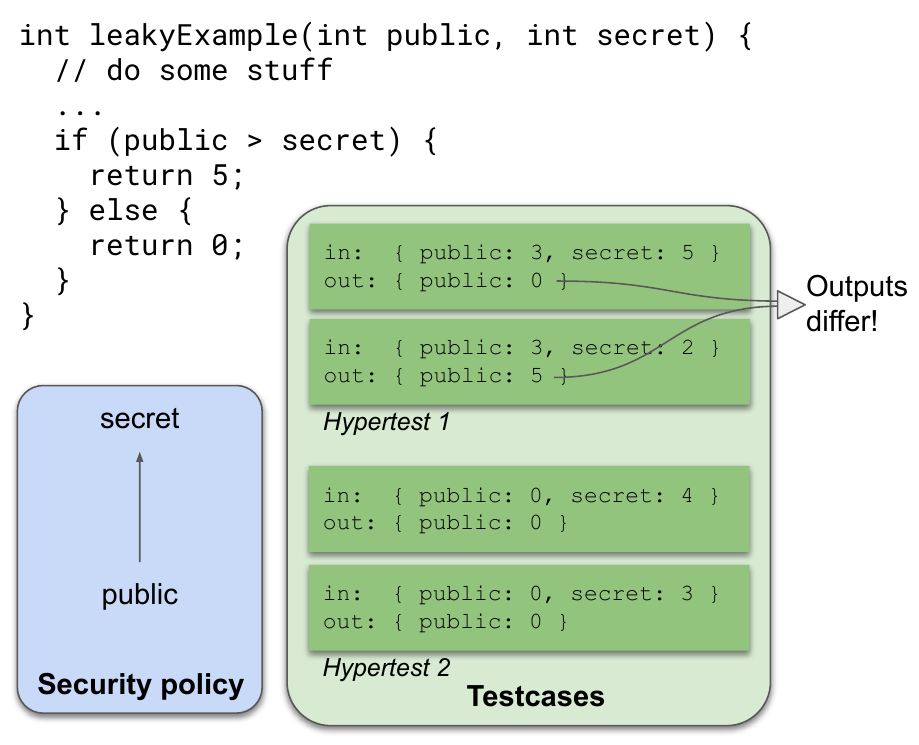}
	\caption{Synthetic code example with an accompanying security policy and two hypertests, one exposing a violation of non-interference and the other failing to do so}
	\label{fig:h-tests}
\end{figure}

The recognition that noninterference is formally a hyperproperty was published in 2008 by Clarkson and Schneider \cite{ClarksonS2008}. 
They distinguished properties of single program executions, such as non-termination, from security properties such as noninterference or bounds on leakage. 
These latter formal properties must necessarily be expressed in terms of sets of executions. 
As explained in Section 2, our noninterference  property is expressed as a universal quantification over \emph{pairs} of program executions.
So a failure of the noninterference property must need a pair of tests, i.e. a hypertest, to witness a fault.
Figure \ref{fig:h-tests} illustrates the hypertest concept.

In Figure \ref{fig:h-tests} a function, \verb|leakyExample| is paired with a security policy where {\tt secret} has a higher classification than {\tt public}.
The input parameters names align with the security policy names, and the return value of \verb|leakyExample| is visible to public.
Hypertest 1 has two inputs, the first has $\langle$public = 3, secret = 5$\rangle$, the second has $\langle$public = 3, secret = 2$\rangle$. 
Note that only the value of \verb|secret| changes between the two inputs. 
However the output values from the function \verb|leakyExample| are different. 
The forgoing is a single hypertest (pair of tests in this case) that exposes a violation of the non-interference property by the program/policy pair. 
Not every hypertest exposes a violation as shown by hypertest 2 in Figure \ref{fig:h-tests} where there is no observable difference in outputs. 
Security properties are further discussed in Section 2.

The OpenSSL cryptography library containing the Heartbleed bug also contained automated tests yet the bug was undetected for over 2 years.
The programming error resulted in a buffer over-read, which exposed secret program memory information to attackers.
This particular class of error is automatically detectable by fuzzing in combination with AddressSanitizer~\cite{serebryany2012addresssanitizer} -- now included with Clang and GCC.
In fact, the OpenSSL version containing Heartbleed is now included as one of the benchmarks in the commonly used fuzzer-test-suite~\cite{fuzzertestsuite}. Security related testing of this kind relies on single tests that detect known, potentially critical error patterns. Such a pattern may or may not correspond to an actual leak, depending on the security policy and other factors.
The strength of hypertesting against a security property is both that leak detection is completely general, i.e.  independent of the nature of the leak, and that there are no uncertainties. Either a hypertest witnesses a violation or it does not.

The expanded capabilities for leak detection offered by hypertesting against a secure flow property can not be done justice by existing benchmarks so we collected a set of 10 C/C++ programs of varying sizes, from 82 to 905,264 lines of code, seven of which contain leaks that have been assigned CVE numbers, into a test bench we call Secure Information Flow Faults (SIFF). 
We have used these SUTs to compare LeakFuzzer against existing state of the art approaches to leak detection. 
Specifically we have compared LeakFuzzer against a model checker, the C Bounded Model Checker (CBMC), and conventional AFL++  augmented in turn with two different sanitisers, MemorySanitizer (MSan) and DataFlowSanitizer (DFSan). 
Unlike CBMC, the mutational engine of AFL++ is nondeterministic so multiple runs were required for comparison purposes. 
Each setup being run 20 times for up to 24-hours, over 400 CPU days of cumulative evaluation was performed.

LeakFuzzer found every leak in every program but not with every campaign, however the majority of the 24 hour LeakFuzzer campaigns did find the sought for leak. In comparison to LeakFuzzer, AFL++,  augmented with either sanitizer, was able to detect leakage in 3/10 SIFF SUTs while CBMC could detect leakage in 2/10 SIFF SUTs. LeakFuzzer tended to use more memory than AFL++ and less than CBMC, which exhausted the available 128GB of memory on one SUT, but there were no issues with memory exhaustion for \leakfuzzer~ in any experiment.

The  contributions of this paper can be summarised as:
\begin{itemize}
\item The creation of \leakfuzzer, the first hyperfuzzer to test against an input-output non-interference property,
\item The creation of the Secure Information Flow Faults benchmark suite (SIFF), a set of varied C/C++ programs for assessing security leak finding tools,
\item A rigorous evaluation study using SIFF that assesses the efficacy and efficiency of \leakfuzzer~ and compares it with three other tools: CBMC, and AFL++ augmented first with DataFlowSanitizer  then with MemorySanitizer.
\end{itemize}

%% file: background.tex
\section{Background}

In this section we elaborate the concepts on which previous information leakage detection tools and \leakfuzzer{} are built: security policies, the noninterference property, hypertesting and greybox fuzzing.

\subsection{The Information Flow Control Problem}

Information outside programs (inputs and outputs) that travels over networks can be encrypted, providing a level of security. 
The interface(s) to programs may need to distinguish between different groups of users based on their privileged access to information, hence the access control problem. 
But implicit within the problem of designing programs with correct access control for different privilege groups is that of maintaining correct information flow control during program operation. 
It wouldn't do to employ encryption and access control correctly but the program operation itself leaks information between the privilege groups. 
All three techniques, encryption, access control, and flow security are required for ``end to end'' security of computing with information. It is this last problem that \leakfuzzer{} addresses. 

Contemporary approaches to information flow control use a policy for the program, variously called a secure flow policy, a non-interference policy, or simply a security policy, depending on the context. 
One of the earliest and best known, developed for the USA military, is the 1973 Bell Lapadula model of access control that provided flow guarantees if the model was followed ~\cite{bell1973secure} .  
Their work formalised security policies and introduced the notion of a security classifications for each piece of information in a program. Denning subsequently recognised that the the Bell and LaPadula principle of ``no read ups and no write downs''  between higher privilege and lower privilege produced an ordering, in fact a partial order, and that this could be extended to a lattice ordering, for completeness ~\cite{denning1976lattice}.  
She then introduced Lattice Based Access Control (LBAC) in which program data containers are mapped to lattice points that represent access control categories.  
The lattice ordering is interpreted as that information can only flow within the program from a lower classification to a higher one. 
To sum up, program users are divided into privilege groups as are data containers (variables etc.) within the program. 
A lattice is constructed with nodes labelled with the privilege groups and the ordering of the lattice expresses the constraints on information flow between them. 
The simplest and most widely used lattice for security policies is the two point or High-Low lattice with the ordering Low $\sqsubseteq$ High although Denning's 1976 observation means arbitrarily complex security policies with many privilege groups are expressible in Lattice Based Access Control. 
In this paper every program discussed or experimented on has a security policy that is a mapping of its data containers to either High or Low in the High-Low lattice.

A program and a security policy being correctly aligned is expressed via a security property. 
The commonly used security property for many program-policy pairs stems from the 1982 paper by Goguen and Meseguer~\cite{gogmes} which introduced the \emph{non-interference principle} described in Section 1. 
This is commonly interpreted as an input-output property in terms of the behaviour of the program and this is what \leakfuzzer~ uses.

\emph{A program $P$ satisfies non-interference if and only if for any pair of low equivalent initial states, the resulting final states from running $P$ with these initial states are also low equivalent}. 

Here, low equivalence of states means that the parts of the two states that are labelled as Low contain the same values. 
Essentially, \leakfuzzer{} is using the non-interference property as a (hyper)test oracle.

As another example, consider a simple program that only takes input from a \emph{High} users, and makes all output available to any user (i.e. to  \emph{Low} users):

\begin{verbatim}
password = read_from_high()
print(password)
\end{verbatim}

As usual our secure flow policy is based on the High-Low lattice. 
The only data container in the program is the variable \verb|password| that is labelled High as it is written by input from a high user. Since the data in \verb|password| is printed it is effectively made available to Low users.
From the source code, it is clear that the \emph{secret} value \verb|password| is copied to the program output visible to \emph{Low}; and thus violates the security policy.
We can witness this violation by providing a pair of inputs that differ only in their \emph{High} part, and consequently differ in their \emph{Low} output.
For this program any pair of inputs are Low equivalent as there is no Low labelled data container, hence a non-interference violation witnessing hypertest could be the pair of \emph{High} inputs ``test'' and ``pass'' that produce \emph{Low} outputs ``test'' and ``pass''. 
The inputs were Low equivalent and the outputs then different.

\subsection{Side-Channel Leakage}

At a high level, side-channel leakage is leakage of information through means other than program output; for example execution time or power consumption.
Being able to detect differences in these observables that are dependent on the High input values allows an attacker to deduce information about those values.
Consider the following program:
\begin{Verbatim}[numbers=left]
guess = read_from_low()
password = read_from_high()
if guess.length != password.length:
    exit
for i in 0..guess.length:
    if guess[i] != password[i]: 
        exit
\end{Verbatim}

Here the security policy again uses the High-Low lattice and here {\tt read\_from\_low()} takes input from Low, and {\tt read\_from\_high()} takes input from High, thus the variables {\tt guess} and {\tt password} contain Low and High information respectively. 
Firstly the length of the input is checked against the length of the actual password, exiting if they are not equal.
The `for' loop in lines 5-7 checks each character of the input \verb|guess| against the actual password, and when a single character does not match the program exits.
Assuming that \emph{Low} is able to observe the runtime of this process, and \verb|read_from_high()| always takes the same amount of time, \emph{Low} can learn information about the degree of correctness of their guess.
The longer the program runtime, the closer their guess is to being correct and thus the more information they learn about the value of \verb|password|.

Notably there is no output from this program, and this is because outputting the result (correct or incorrect) would in itself leak information about the value of {\tt password}.

\subsection{Self-Composition}
\label{sec:self-comp}

Self-composition \cite{barthe2011secure}, as the name suggests, involves composing a program with itself in such a way as to provide the two executions with the same Low inputs but differing High (secret) inputs; the Low outputs can then be compared to check that they match (indicating non-interference).
This approach is used by the side-channel fuzzers discussed in the later related work, as well as the model checking approach that we compare against in the evaluation of \leakfuzzer{}.
Self-composition is useful as it can convert a hypertest into a single test.
See the following example where the input parameter {\tt public} is classified as Low and {\tt secret} is High:

\begin{verbatim}
int isLarge(int public, int secret) {
    // do some stuff here, does not matter what
    ...
    if (secret > 2) return 1; else return 0;
}
\end{verbatim}

\noindent To implement the self-composition approach we would create the following wrapper function that compares the output of \verb|isLarge| with two different \verb|secret| values:

\begin{verbatim}
int selfComposedIsLarge(int pub, int sec1, int sec2) {
    assert(isLarge(pub, sec1) == isLarge(pub, sec2));
}
\end{verbatim}

Any testcase for the wrapper function that causes the assertion to fail here, for example {\tt $\langle$ pub = 0, sec1 = 1, sec2 = 3 $\rangle$}, is a witness to the information leak in \verb|isLarge|.
Essentially self-composition allows a hypertest to be performed with a single testcase (given a wrapper function), thus allowing it to be used by techniques that do not natively support hypertesting.

\subsection{Fuzzing}

Fuzzing is program testing technique.
Generally it is considered a system testing technique, but could be applied to smaller units, and is generally applied with the intention of discovering security vulnerabilities.
The term fuzzing dates back to at least 1990 \cite{miller1990empirical}.
At its core, it is closely related to random testing and the process is orchestrated by a tool called a fuzzer.
The fuzzer is responsible for generating test cases, executing them and watching for crashes within the program.
The term \emph{fuzzing campaign} is used to refer to the fuzzing process from start to finish.
These fuzzing campaigns run for a long time, regularly at least 24 hours~\cite{klees2018evaluating}, but often much longer~\cite{moroz_2019}.

Like software testing in general, there are 2 opposing top-level approaches to building a fuzzer: a black-box approach which has no knowledge of the SUT's (System Under Test) internal structure and state, whereas a white-box approach uses program analysis to improve exploration of program behaviours. 
An example black-box fuzzer is Zzuf~\cite{hocevar_2007}, the key advantage of the black-box approach is the sheer rate at which test cases can be generated and ran. 
An example white-box fuzzer is SAGE~\cite{godefroid2012sage} which leverages symbolic execution to allow it to systematically test different execution paths; this process suffers from the typical limitations of symbolic execution, which limits scalability.
Indeed some of the approaches based around fuzzing are incredibly complicated, take for example DriFuzz~\cite{shendrifuzz}, which uses concolic execution to figure out how to successfully initialise hardware device drivers; these drivers are executed inside of a full system emulation (using QEMU~\cite{bellard2005qemu}).

There is one further class of fuzzer, lying between black-box and white-box classes is the aptly named grey-box class, one of the most well known is American Fuzzy Lop~\cite{zalewski} (AFL) and its updated relative AFL++~\cite{fioraldi2020afl++}.
AFL instruments the binary for the SUT in such a way as to receive coverage information detailing which branches were run in an execution; testcases are mutated and mutations run, with the fuzzer storing any that lead to new coverage in a queue for further mutation. 
Using this queue-based mutation approach, AFL is able to explore a wide range of the SUT's behaviours without the program size and complexity constraints of a white-box fuzzer.
It is worth noting that AFL includes a compiler wrapper that generates the necessary instrumentation (but can compile only C/C++), but it is also capable of fuzzing uninstrumented binaries using its built in QEMU mode.
There are many extensions to AFL to support Python, Rust, JavaScript, Java, Swift, OCaml and .net~\cite{aflBasedFuzzers}.

\subsection{Sanitizers}

Whilst fuzzing primarily aims to detect crashes in programs, modifications can be made to the program under test to allow for a broader range of errors to be detected.
One of the simplest ways to achieve this is to add assert statements that trigger a crash whenever a bad state is reached.

There is a family of these error detectors included in the LLVM compiler infrastructure referred to as \emph{sanitizers}~\cite{sanitizers}.
As such, they are compiled into the program under test and doing this is usually as simple as adding a compilation flag for C and C++ programs.
When an error is detected by the sanitizer at runtime, the program is crashed and thus the error detected by the fuzzer.
The sanitizer dumps useful information such as stack trace and other details before exiting, which aids in the debugging process later on.

AddressSanitizer detects addressability related memory issues such as use after free, buffer overflows and use after scope; it also includes LeakSanitizer which detects memory leaks.
ThreadSanitizer detects data races and deadlocks, MemorySanitizer detects use of uninitialised memory and UndefinedBehaviourSanitizer detects undefined behaviour such as integer overflow and bitwise shifts out of bounds.
DataFlowSanitizer is an implementation of dynamic taint analysis, which requires modifying the test program to insert API calls to label data and check for label propagation.

\section{Related Work}
\subsection{Fuzzing Applied to Side-Channel Leakage}

Fuzzing has been applied to side-channel leakage detection, firstly in 2019 by a tool called DifFuzz~\cite{nilizadeh2019diffuzz}.
DifFuzz works on Java programs, and makes use of self-composition.
It executes the target program with two differing secret inputs, and counts the number of JVM (Java virtual machine) bytecode instructions executed in each case; if these differ then it is flagged as a timing leak.
A 2021 follow on paper sharing an author with DifFuzz is QFuzz~\cite{noller2021qfuzz}, which also provides an estimate on the quantity of information leakage through the timing side-channel.

A 2020 paper describes a fuzzing tool---ct-fuzz~\cite{he2020ct}---for detecting side-channel leakage in C and C++ programs.
Like DifFuzz, it uses self-composition, but has a more complex model for estimating runtime: firstly it checks that both executions follow the exact same path, and additionally a CPU cache model is used to determine whether any differences in cache misses exist.

Another 2020 paper describes a fuzzing-based approach for detecting JIT-induced side-channels in the JVM~\cite{brennan2020jvm}.
Here JIT refers to just-in-time compilation, which is used to compile heavily used sections of code at runtime, speeding them up.
Statistical methods are used to determine the amount of secret information that can be learned from the differences in execution time caused by the JIT compilation.

Note that three of these papers use models to \emph{estimate} runtime differences, which can vary significantly between hardware setups.
The final paper overcomes this by using real execution times combined with statistics, but would still need each program to be tested on the target hardware setup ideally.
Additionally, none of these tools are able to detect output-based information leaks, as \leakfuzzer{} does, as the detection of side-channel leakage is a fundamentally different problem.

\subsection{Quantified Information Flow}

There is a body of work on static or dynamic analysis for measuring the size of leaks that began with Clark, Hunt and Malacaria \cite{clark2007static}. Detecting a leak and measuring a leak are two different things but depending on the algorithm for measuring it, they can overlap. Heusser's work falls into this category ~\cite{heusser2010quantifying} as does that of Biondi mentioned above ~\cite{biondi2018scalable} as well as the the tool LeakiEst ~\cite{LeakiEst} . There are other approaches too numerous to mention here but to date these approaches either are static and don't scale well, like CBMC, or are dynamic but don't automatically generate inputs, like LeakiEst or LeakWatch  ~\cite{LeakWatch}, or are dynamic and automatically generate inputs but are limited to pseudo-random number sampling.

\subsection{Constraint Solving Approaches to Quantifying Information Flow}

An alternative approach to \leakfuzzer{} is constraint based model checking, a formal verification technique.
A 2010 paper by Heusser and Malacaria~\cite{heusser2010quantifying} used the CBMC (Bounded Model Checker for C) tool to provide a lower bound on the quantity of information leakage in C programs, including evaluation on six real world programs (four of which contained reported CVEs).
The method described there iterates over guesses at the channel capacity of a leak so that an upper bound on it can eventually found.  

Integral to the process is that it discovers by constraint solving whether two or more distinct outputs can be produced as a counter example to a channel capacity assertion by providing the program with inputs that do not differ in their non-secret part.  

A 2012 paper by Phan, Malacaria et al.~\cite{phan2012symbolic} used symbolic execution to quantify information flow, in this case using Java Pathfinder (JPF), which as the name suggests operates on Java bytecode programs.
This technique is similar to that explained above, but was evaluated only on fabricated example programs.

It is acknowledged that there are issues with scaling these verification techniques to work with larger programs.
This is mentioned in a 2018 paper, and a solution proposed; this being to approximate the model to reduce solving complexity~\cite{biondi2018scalable}.

\subsection{Hypertesting}

Recent papers on HyperGI, a technique that detect leaks and repairs them also use hypertests~\cite{MesecanBCCP2021,MesecanBCCP2022}. 
However, they do not have a method for automatic hypertest generation.

%% file: technique.tex
\section{\leakfuzzer}

Formal verification techniques for discovering program leakage tend to suffer from an inability to scale, unlike software validation (testing) techniques.
However many validation-based approaches do not solve the problem of testcase generation, which is a significant issue for programs accepting complex inputs.
In comparison, fuzzing is a validation technique that is noted for its ability to handle testcase generation, achieve good program exploration and scale well to large programs.
%Being a testing technique it cannot confirm the absence of errors. \ingolf{maybe cut this sentence?}
Here we present \leakfuzzer{}, a fuzzing-based tool developed with the goal of providing automated, scalable information leakage detection.

\begin{figure}
	\includegraphics[width=\columnwidth]{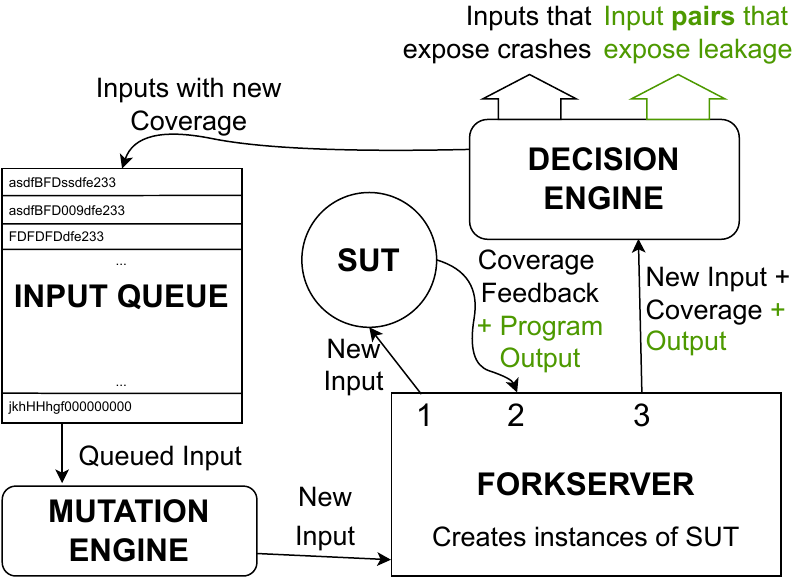}
	\caption{High-level architectural overview of a grey-box fuzzer (black), with additions for \leakfuzzer{} (green).}
	\label{fig:architecture}
\end{figure}

\leakfuzzer{} is based on the popular AFL++ fuzzer~\cite{fioraldi2020afl++}, and as such contains all of the basic components of a standard grey-box fuzzer as used for finding regular program errors; these are shown in black in figure \ref{fig:architecture}.
There is an \emph{input queue} that stores `interesting' inputs; in \leakfuzzer's case this is unmodified from AFL++, whereby an interesting input is one that previously achieved unseen program coverage and therefore should be mutated further to generate new inputs by the \emph{mutation engine}.
The \emph{forkserver} creates instances of the system under test (SUT), feeds them the mutated inputs and collects coverage information.
Finally, the \emph{decision engine} receives the coverage information and decides whether this input is `interesting' and thus should be stored in the \emph{input queue}.

We have made a number of architectural modifications and additions in order to detect a new class of error: information leakage.
The conceptual changes are shown in green in figure \ref{fig:architecture}. 
Firstly, program output is fed back through the forkserver to the decision engine.
Secondly the decision engine has been modified to process program output and store input pairs that expose information leakage.
Finally, the mutation engine has been modified as described in section~\ref{sec:generatingPublicPrivatePairs}.

\leakfuzzer{} is able to detect a superset of the errors that AFL++ can.
The hypertesting approach used in \leakfuzzer{} can be generalised to test for any hyperproperty, and we call a fuzzer that implements such an approach a hyperfuzzer.
Note that in the following sections a simple security policy is assumed, with two classifications \emph{secret} and \emph{public}, where \emph{secret} is of a higher classification than \emph{public}.

\subsection{Hypertesting Approach}
\label{sec:hashmap}

By using a hashmap and storing outputs \leakfuzzer~ only needs to run each input once. This contrasts with the efficiency of methods that use self-composition (discussed in section \ref{sec:self-comp}) to achieve hypertests. Suppose that there are $N$ distinct High labelled values in the test set. For each possible Low labelled input, self composition must in the worst case explore a space of $C^N_2$ explicit pairs, i.e. $O(N^2)$, as opposed to \leakfuzzer's exploration of a space of at most $O(N)$.

In order to illustrate this, let's consider this simple program which takes no public inputs, one secret input---a 2-bit unsigned integer \verb|int| (taking values between 0 and 3 inclusive)---and returns a value that is visible to public:

\begin{verbatim}
int isLarge(int secret) {
  if (secret > 2) return 1; else return 0;
}
\end{verbatim}

\noindent Using the self-composition approach we could create the following wrapper function:

\begin{verbatim}
int selfComposedIsLarge(int secret1, int secret2) {
  assert(isLarge(secret1) == isLarge(secret2));
}
\end{verbatim}

The six possible non-repeating inputs to \verb|selfComposedIsLarge| are \verb|{0, 1}, {0, 2}, {0, 3}, {1, 2}, {1, 3}, {2, 3}|.
Of these, \verb|{0, 3}, {1, 3}| and \verb|{2, 3}| cause the assertion to fail.
This means that three out of the six ($\frac{1}{2}$) hypertests expose the leak.
Therefore if we were to sample hypertest pairs uniformly, we would expect that two $(1 \div \frac{1}{2})$ hypertests need evaluation before we detect the leak. 
This means running \verb|isLarge| four times; but it could potentially be as high as eight times in the worst case (that is, the three hypertest pairs that do not detect leaks are chosen first).

The approach taken by \leakfuzzer{} instead requires each input to be tested no more than once by caching the output so that it can be compared against in future without needing to be rerun.
Caching outputs in full would require too much memory, hence we store 64-bit hashes of program outputs alongside 64-bit hashes of the corresponding program public and secret inputs.
The XXH64 hash \cite{xxhash} is used due to it's extremely high throughput.
Not only can we store and compare outputs for programs with large output spaces, but the comparisons can be done in a single machine code instruction due to the size of the stored hashes.
This sacrifices some memory efficiency compared to naive self-composition---which requires no memory of past inputs and outputs---in exchange for improved time efficiency as each input can be executed just once.

A detailed pseudocode listing of the leak detection algorithm used by \leakfuzzer{} can be found in figure \ref{fig:pseudocode}, and line numbers in the following prose reference this figure.
A hash table is held in memory by \leakfuzzer{}, with the public input hashes used as keys, that map to values consisting of a structure containing secret input hashes, output hashes and other properties (lines 6--11).
Hashes are capable of colliding, however for a 64-bit hash, collisions are unlikely to effect a fuzzing campaign with 1 billion ($\approx2^{30}$) or so inputs significantly.
Using this hash table we can detect differences in output within pairs of testcases where public inputs are identical, and secret inputs differ. This is sufficient to infer that information leakage does occur.

This does not ensure that those output differences are not caused by some form of non-determinism as opposed to being deterministically influenced by the secret input.
To mitigate this, \leakfuzzer{} stores any testcases that have resulted in a different output to other testcases with an identical public input, reruns them 100 times and directly compares the output buffers byte-by-byte, with any testcases that give flaky (inconsistent) outputs being discarded (lines 26--32).
Any testcases that pass this \emph{flakiness test} are stored in full in the hash table value's structure (line 38), and once a pair of non-flaky, output differing testcases has been found for any public input hash then both are reported as a leaking hypertest pair and output to file (lines 40--44).

\begin{figure}
\begin{footnotesize}
\begin{Verbatim}[numbers=left]
struct Input {
  ByteArray public;
  ByteArray secret;
};

struct HashValue {
  int64 secretInputHashes[];
  int64 publicOutputHashes[];

  ByteArray secretInputsFull[];
};

HashTable<int64, HashValue> dict;

// Keep going indefinitely until fuzzing campaign halted
while true:
    Input generatedInput = generateNewInput();
    ByteArray publicOutput = targetProgram.run(generatedInput);

    int64 publicInputHash  = hash(generatedInput.public);
    int64 secretInputHash  = hash(generatedInput.secret);
    int64 publicOutputHash = hash(publicOutput);

    if (HashValue value = dict.get(publicInputHash)) != null:
        if not value.publicOutputHashes.contains(publicOutputHash):
            is_flaky = false;
            for _ in 0..100:
                if targetProgram.run(generatedInput) != publicOutput:
                    is_flaky = true;
                    break;
            if is_flaky:
                continue;  // discard this flaky input

            value.secretInputHashes.append(secretInputHash);
            value.publicOutputHashes.append(publicOutputHash);

            // Only store the full input if a suspected leak is found
            value.secretInputsFull.append(generatedInput.secret);

            cnt = secretInputsFull.length;
            if cnt % 2 == 0:
                outputHypertestToFile(generatedInput.public,
                                      value.secretInputsFull[cnt-2],
                                      value.secretInputsFull[cnt-1]);

    else:  // publicInputHash not found in dict
        HashValue newValue = {
            secretInputHashes = [secretInputHash],
            publicOutputHashes = [publicOutputHash],
            secretInputsFull = []
        };
        dict.set(publicInputHash, newValue);
\end{Verbatim}
\end{footnotesize}

\caption{Pseudocode algorithm describing the hypertesting approach used by \leakfuzzer{}.}
\label{fig:pseudocode}
\end{figure}

%\vspace*{-0.2cm}
\subsection{Generating \{Public, Secret\} Input Pairs} \label{sec:generatingPublicPrivatePairs}

The basis for \leakfuzzer{}, AFL++, has no awareness of the internal structure of inputs by default, and as such generates only raw byte arrays.
It is up to the \emph{fuzzing harness} to parse these byte arrays into a format that can be used by the program being tested, this is at its core a similar process to deserialisation.

A naive approach to deserialising a fuzzer generated input into public and secret parts could be taken if either the public or secret parts were of fixed length $x$.
One could simply read in the first $min(\text{input.length}, x)$ bytes needed to fill the fixed length part, and then read the remaining $max(\text{input.length} - x, 0)$ bytes into the other part.
If both parts are of fixed length $x$ and $y$ bytes, then the $min(\text{input.length} - x - y, 0)$ bytes beyond those required can simply be discarded.

In the case that both input parts can be of variable length, then the fuzzer generated input could be split in half, with the first part used as input for Low and the latter used as input for High.
Alternatively, the first 1 (or more) byte(s) could be interpreted as an integer $x$ indicating the length of the secret part of the input; the first $x$ bytes after the length indicator would be read into the secret input, and the remaining bytes (if any) read into the public input.

These simple approaches require no adaptations to the fuzzer, however the fuzzer has no awareness of which parts of the generated input correspond to the public and secret inputs.
In order to detect differences in output caused by information leakage, inputs must only differ in their secret parts.
Without awareness of which parts of the input are secret, the fuzzer cannot be certain of whether any observed differences are due to the public or secret parts of the generated inputs.

In order to overcome this difficulty, \leakfuzzer{} is adapted to have oversight of public and secret parts of generated inputs.
This separation of input segments has several advantages.
First and foremost, it allows \leakfuzzer{} to determine whether output differences are caused by information leakage (assuming a fully deterministic program).
Secondly, it allows for specific targeting of mutations on either part of the input (as selected from the fuzzer \emph{input queue}).

\begin{figure}
	\centering
	\includegraphics[width=\columnwidth]{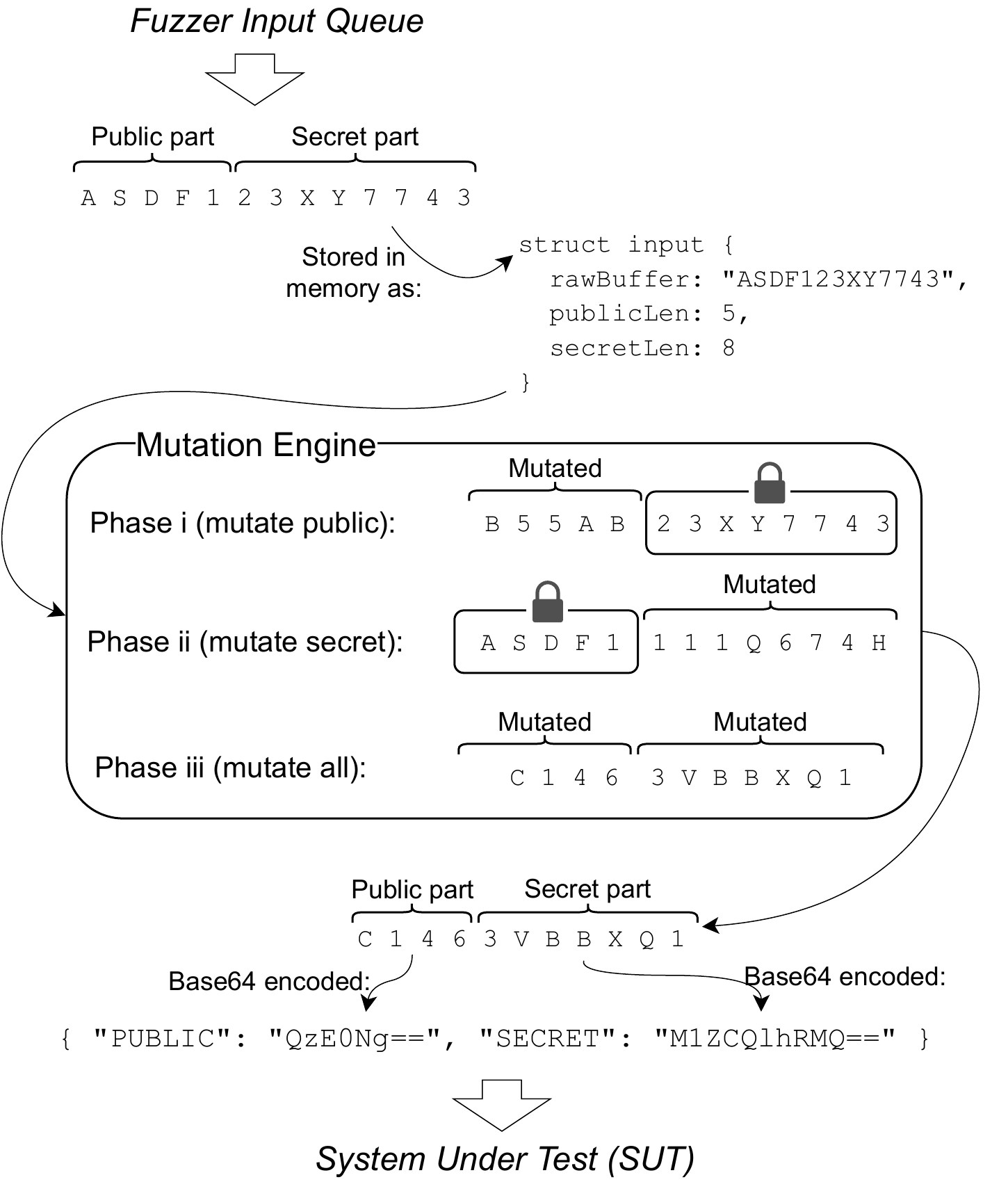}
	\caption{Block diagram showing the internal and external representations of inputs in \leakfuzzer{}.}
	\label{fig:mutator}
\end{figure}

We have discussed why a hypertest exposing an information leak will have identical public inputs, but differing secret inputs, so we may wish to schedule a period that mutates the secret part of the input exclusively. 
Figure~\ref{fig:mutator} shows the memory structure of the stored inputs, a visualisation of the mutation phases and finally the encoded input exactly as it will be provided to the SUT (note that fuzzing harnesses must decode this back into byte arrays, and a utility function to do this is included with the \leakfuzzer{} source).

\leakfuzzer{} has three mutation phases targeting: the public part of the input, the secret part of the input, and the entire input.
All three strategies are used because any part of the program input could decide whether a particular path is chosen, and branch conditions may depend on some aspect of both the public \emph{and} secret parts of the input.
Note that the lengths of both the public and secret parts can be altered by the mutations, and this is accounted for in the fuzzer logic.
The set of available mutations that can be applied are unaltered from AFL++.

\subsection{Handling Invalid Memory Reads}
\label{sec:uninit}

The following subsections detail the types of memory misusage that can lead to information leakage, and how \leakfuzzer{} can detect this.

\subsubsection{Uninitialised Memory Reads}

\leakfuzzer{} targets C and C++ programs, both of which require manual memory management.
One of the consequences of this is that variables are not automatically initialised at declaration.
This uninitialised variable (section of memory) then contains the values that previously occupied this section of memory; this information potentially contains secrets.
One would assume that this type of error is rare, however when using \verb|struct|s this mistake is much easier to make.
It is common to declare a struct instance, and then populate the member values manually.
This approach does avoid writing to each member twice (first to zero it, then secondly to set it to the final value), however it also means that if the developer forgets to set a member then this is left uninitialised.
If this uninitialised value is then output from the program, it can leak secret information.

\subsubsection{Out of Bounds Memory Reads}

Closely related to uninitialised memory usage is out of bounds memory usage, as both can cause information leakage via the reading of values that were not assigned to in the current scope.
Heartbleed was caused by data being copied from out of bounds accesses to a buffer to program output.
An additional source of these errors that can exist in C and C++ is the use of \verb|memcpy| with structs.
This is due to \emph{memory alignment}, which is done for performance reasons.
Take, for example, the following struct:

\begin{verbatim}
struct instruction_t {
  char direction;
  int  distance
};
\end{verbatim}

A tightly packed structure on a 64-bit machine (with 32-bit \verb|int|s as per gcc) would take up only 5 bytes: 1 byte for \verb|direction| followed by 4 bytes for \verb|distance|.
However, assuming a preference for 4-byte memory alignment, this struct would actually take up 8 bytes: 1 byte for \verb|direction|, 3 bytes of padding to reach a 4-byte boundary, and 4 bytes for \verb|distance|.
Naively serialising this struct to program output by using \verb|memcpy|, the 3 bytes of uninitialised padding would also be output, potentially leaking secret information.

\subsubsection{Solution}

The \emph{fuzzing harness} that is used to convert the raw byte-string generated by the fuzzer into a form of input accepted by the program often runs only a single input through a slice of the system.
As memory pages are zeroed by the OS before being passed to the program process, it is common that early uninitialised memory reads access these zeroed regions.
The longer that the program runs, the more common it is that memory is reused; and this is particularly the case in long running interactive programs such as web servers and databases.
It is therefore possible that invalid memory reads that lead to information leakage to output can go unnoticed in these fuzzing harnesses, but would affect live long-running processes.
Additionally, the output would change depending on the previous value stored in that area of memory and thus may appear to exhibit a form of non-determinism.

In order to allow for the detection of these potential information leaks, \leakfuzzer{} uses a technique to set internal program memory to a pattern of consistent non-zero values.
To do this, a portion of the generated \emph{secret} input is used as a seed for a pseudorandom number generator, and a sequence of bytes is generated using this (in the current implementation, 8 bytes).
This sequence of bytes is then replicated, until it fills the process stack memory, by a function within the initialisation phase of the fuzzing harness.
Heap memory initialisation is handled by a provided wrapper for the \verb|malloc| function, this wrapper first allocates the required memory and then fills it with the repeated pseudorandom byte sequence.

By seeding the pseudorandom sequence from the \emph{secret} input, it becomes possible for \leakfuzzer{} to generate hypertests that can produce \emph{deterministic outputs} due to invalid memory reads and thus expose the information leakage.

%% file: evaluation.tex
\section{Evaluation and Results}

Having described \leakfuzzer{} and how it works, in this section we evaluate how well it performs against information leakage faults in the benchmark software suite we have assembled. We assess the usual questions of efficacy and efficiency. We then evaluate how well its performance compares to a representative sample of available tools that can be used to discover leaks.

\subsection{Research Questions}
\noindent
\textbf{RQ1:} How many known information leaks in our SUTs are discovered by \leakfuzzer{}? \\
We seek to answer efficacy permissively, by the percentage of SUTs in which the known error is found in at least one of the 20 24-hour runs.\\
\textbf{RQ2:} In what proportion of runs does \leakfuzzer{} detect leaks within a standard 24-hour fuzzing budget? \\
Here we seek to determine the efficiency of \leakfuzzer{} by considering the mean time to discovery and the proportion of runs that detect leakage within a standard fuzzing campaign budget of 24 hours. \\
\textbf{RQ3:} How efficacious and efficient is \leakfuzzer{} in comparison with existing practical approaches that can detect information leakage?\\
We compare with three tools: MemorySanitizer, DataFlowSanitizer, and the C Bounded Model Checker (CBMC).
We compare the results using the same metrics as used in RQ1 and RQ2 for all three tools.
Additionally we look at measurements of memory usage.

First we discuss the creation of the benchmark suite.

\subsection{Secure Information Flow Faults (SIFF) Benchmark Suite} 

No benchmark suite for information leaks in C and C++ was available to our knowledge. 
Hence we created SIFF, a secure information flow faults repository that contains programs together with fuzzing harnesses and leak exposing hypertests. Information flow control issues (information leaks) are difficult to find in known issue databases due to the language used to describe them. 
The word \emph{leak} is overloaded and could refer to memory leaks, resource leaks or confidential information leaks, with the former two being more common in C and C++ programs. 
In total we have included 10 distinct programs, of which 7 are taken from confirmed CVE reports and others are example programs taken from existing work. Implicit and explicit flows are represented in the program suite as well as memory management and design issues. It also has examples of different sized SUTs and a mixture of kernel and user space programs. In addition, the SUTs have variety in input types including  byte strings, human readable formats and SQL queries. The programs taken from CVE reports span errors from 2007 to 2022.  The repository is  publicly available for reuse in future research.

A fuzzing harness has been constructed for each individual program. 
The OpenSSL-1.0.1f fuzzing harness was modified from the version contained in fuzzer-test-suite \cite{fuzzertestsuite}. 
Additionally, when testing using AFL++ based systems, the PostgreSQL program uses the AFL++ Grammar Mutator \cite{aflgrammarmutator} plugin to generate SQL statements from a provided grammar. 
This is because the standard mutation engine struggles to generate valid statements even when provided many seeds.

\begin{table}[H]
	\caption{Program details for members of the Secure Information Flow Faults benchmark suite.  Note that Leak Type is abbreviated using the key: E: explicit flow, I: implicit flow, M: Memory issue, PD: Program design issue.}
	\vspace{-0.3cm}
	\label{table:BenchmarkPrograms} 
	\begin{small}
	\begin{center}
	\begin{tabular}{lllllll} \toprule
	Program Name              & Leak Type        & LoC           & CVE Number     & Source            \\ \midrule
	appletalk                        & E + M                       & 110           & CVE-2009-3002  & \cite{heusser2010quantifying, klebanov2013sat} \\
	Banking      &  I + PD  & 150           & -              & \cite{hamann2018uniform}            \\
	cpuset                    & E + PD               & 82            & CVE-2007-2875  & \cite{heusser2010quantifying} \\
	NetworkManager            & E + PD               & 18,185         & CVE-2011-1943  & -                 \\
	OpenSSL-1.0.1f  & E + M           & 279,466        & CVE-2014-0160  & \cite{fuzzertestsuite} \\
	Password Check  & I + PD    & 117           & -              & \cite{hamann2018uniform}            \\
	PostgreSQL                & E + PD               & 905,264        & CVE-2021-3393  & -                 \\
	RDS                       & E + M               & 94,248         & CVE-2019-16714 & -                 \\
	Reviewers       & I + PD    & 145           & -              & \cite{hamann2018uniform}            \\
	sr9700\_rx\_fixup             & E + M               & 439           & CVE-2022-26966 & -                \\ \bottomrule
	\end{tabular}
	\end{center}
	\end{small}
    \vspace*{-0.4cm}
\end{table}

All programs are written in C with the exceptions of Banking, Password Check and Reviewers; these were originally written in Java and manually converted to C++.

\paragraph{Program Design Errors:}
Of the 10 programs, six contain information leaks caused by program design error, where a flow logic error causes the leakage.

\paragraph{Memory Mismanagement Errors:}
Four of the programs contain memory related errors and all are detected by \leakfuzzer{}.
The first of these is appletalk, this is a simplified version of the original bug taken from a previous paper on measuring information leakage \cite{heusser2010quantifying}. 
This leaks stack memory from within the linux kernel caused by failure to initialise all members of a struct before returning the value to userspace.
OpenSSL-1.0.1f contains the infamous Heartbleed bug which leaked information by outputting internal program heap memory due to a buffer over-read.
RDS (reliable datagram sockets) is also taken from the linux kernel, with a stack memory leak again caused by failure to initialise all struct members.
Finally, sr9700\_rx\_fixup is taken from a linux network driver that leaks heap memory due to buffer over-read.

\subsection*{Testing Environment}

All tests were run inside Docker containers based on the Ubuntu 20.04 distribution of Linux.
Fuzzing experiments were run with 10 fuzzing campaigns in parallel, each bound to a single CPU core.
Model checking (CBMC) experiments were not run in parallel due to RAM space being a common bottleneck.
Tests were run on dedicated servers each equipped with 2 x Intel Xeon E5-2620 v2 processors, making for a total of 12 cores (24 threads) at 2.10GHz, and 128GB RAM. As per is common in the evaluation of fuzzers, each benchmark was fuzzed for 24 hours, though this time length does not produce definitive runs on large programs \cite{klees2018evaluating}.

Each experimental setup/run was repeated 20 times.  
The nature of the mutational engine on the inputs is nondeterministic so different results may occur in different 24 hour runs. 

\section{Results}

\vspace*{-0.25cm}
\begin{table}[H]
    \caption{Results for \leakfuzzer{} on the SIFF benchmark suite.}
    \label{table:leakfuzzerRuntime}
    \vspace*{-0.3cm}
    \begin{small}
		\begin{tabular}{llp{1.5cm}p{1.5cm}} \toprule
			\multirow{2}{*}{SUT} & \multirowcell{2}{\% runs leak\\ detected} & \multicolumn{2}{l}{Leak detected time (s)} \\
			&                                          & Mean                 & Std. Dev.            \\ \midrule
			appletalk                                  & 100                        & 175.23                       & 250.49                         \\		
			Banking						               & 100                        & 1.48                         & 1.25                            \\
			cpuset                                     & 95                         & 1.16                         & 0.41                            \\
			NetworkManager                             & 100                        & 228.01          & 141.43                          \\
			OpenSSL-1.0.1f                             & 75                        & 40,141.11           & 21,129.60                       \\
			Password Check                             & 100                        & 483.65           & 1,575.00                         \\
			PostgreSQL                                 & 55                         & 32,870.09           & 19,788.94                        \\
			RDS                                        & 100                        & 3,223.85                    & 8,910.57                          \\                          
			Reviewers                                  & 100                        & 82.53                    & 87.38                          \\
			sr9700\_rx\_fixup                              & 40                         & 16,530.35          & 27,442.28                        \\ \bottomrule
		\end{tabular}
    \end{small}
\end{table}
\vspace*{-0.1cm}

{\bf RQ1: How many known information leaks in the set of benchmarks are discovered by \leakfuzzer?}

For all 10 programs, information leakage was detected in at least 40\% of the runs.
This includes all leak types, in particular five of the six program design errors were detected in 95\% or more runs.
Discussion on the cause of variability follows in RQ2.

\noindent
\fbox{
\begin{minipage}{0.99\columnwidth}
\leakfuzzer~finds every leak in every program, but not in every run.
\end{minipage}
}

\vspace*{0.2cm}
\noindent
{\bf RQ2: In what proportion of runs does \leakfuzzer{} detect leaks within a standard 24-hour fuzzing budget?}

Each run used the same set of initial seeds. As can be seen in table \ref{table:leakfuzzerRuntime}, even for the same program, runtime before detecting the first leak varied considerably due to non-deterministic input queue and mutation selection strategies used in AFL++.

The two most complex programs as measured by lines of code -- OpenSSL-1.0.1f and PostgreSQL -- had the longest discovery times, as would be expected with an exploratory approach such as fuzzing.
\leakfuzzer{} took a surprisingly long time to discover leakage in sr9700\_rx\_fixup despite containing relatively few lines of code due to the partition of the input space that triggers the leak being very small.
In this particular case there were 28 edges that could be covered, and in 19 of the 20 runs 17 edges were covered; only 12 were covered in the other run.
As might be expected, the information leak was not detected in the run that produced less coverage.
As the other runs demonstrate, there is not a one-to-one relationship between coverage and the discovery of leakage.
We see a similar lack of correlation between coverage and leak discovery for both OpenSSL-1.0.1f and PostgreSQL.

\vspace*{0.1cm}
\noindent
\fbox{
	\begin{minipage}{0.99\columnwidth}
	It is likely that all fuzzing campaigns would have discovered leakage if left to run for long enough; however within the 24-hour time limit imposed here the majority still did.
	\end{minipage}
}

\vspace*{0.2cm}
\noindent
{\bf RQ3: How does \leakfuzzer{} compare with existing approaches that could be used to detect information leakage?}

Both MemorySanitizer (MSan) and DataFlowSanitizer (DFSan) were evaluated using a similar fuzzing harness and input seeds.
These setups were fuzzed by an unmodified version of AFL++ for 24-hours each to provide a fair comparison against \leakfuzzer.
Those programs contining a leak caused by a memory issue were evaluated with MSan, and those containing a program design issue with DFSan.

\begin{table}[H] 
    \caption{Results for AFL++ in combination with the two sanitizers on the SIFF benchmark suite.}
    \vspace*{-0.2cm}
    \begin{small}
		\begin{tabular}{llp{1.5cm}p{1.0cm}p{1.0cm}} \toprule
			\multirow{2}{*}{SUT} & \multirow{2}{*}{Sanitizer} & \multirowcell{2}{\% runs leak\\detected} & \multicolumn{2}{l}{Leak detected time (s)} \\
			&                            &                                        & Mean                 & Std. Dev             \\
			\midrule
			appletalk            & MSan                       & 0                                   & -                    & -                    \\
			banking              & DFSan                      & 0                                   & -                    & -                    \\
			cpuset               & DFSan                      & 100                                 & 0.12                 & 0.11                 \\
			NetworkManager       & -                          & -                                      & -                    & -                    \\
			OpenSSL-1.0.1f       & MSan                       & 100                                  & 2377.47              & 2732.26              \\
			Password Check       & DFSan                      & 0                                   & -                    & -                    \\
			PostgreSQL           & DFSan                      & 0                                   & -                    & -                    \\
			RDS                  & MSan                       & 100                                 & 548.71               & 152.25               \\
			Reviewers            & DFSan                      & 0                                   & -                    & -                    \\
			sr9700\_rx\_fixup    & MSan                       & 0                                   & -                    & -                    \\ \bottomrule
		\end{tabular}
    \end{small}
\end{table}

\noindent
{\bf Memory Sanitizer (MSan)}

As one of the LLVM sanitizers, MSan can be used in combination with fuzzing to detect uninitialised memory usages.
For this reason, it may be able to detect a proportion of the information leaks caused by memory mismanagement errors.
It is worth noting that the presence of uninitialised memory use does not necessarily imply that the program output will be affected, thus there may not be information leakage caused by many errors detected by MSan.
In order to determine whether a crash detected by MSan `detected' an information leak, the stack trace for each discovered error was manually inspected.
If fixing said error would eliminate the leakage then the leak was considered to have been `detected' in that run.
Due to repeats sharing the same executable, each unique crash (i.e. those sharing a stack trace) only needed inspecting once.

The four programs containing leaks caused by memory mismanagement -- appletalk, OpenSSL-1.0.1f, RDS and sr9700\_rx\_fixup -- were evaluated with MSan, as it can potentially detect these.
Of this subset, errors related to the known information leak were discovered in OpenSSL-1.0.1f and RDS by MSan.
It is likely that the error in OpenSSL-1.0.1f was discovered much more quickly by MSan than \leakfuzzer{} as many inputs that trigger the buffer over-read do not reach far enough in memory to result in leakage.
\leakfuzzer{} instead requires that there are discernable differences in output so must find those inputs that trigger this behaviour.
The error in RDS was also detected more quickly by MSan, though not by as significant a margin.
\\

\vspace*{-0.2cm}
\noindent
{\bf DataFlowSanitizer (DFSan)} \nopagebreak

As it provides an implementation of taint analysis, DFSan is more closely related to information flow control than MSan.
Where MSan required no modification to the fuzzing harnesses, for DFSan we added labels to the secret inputs and asserted that this label did not propagate to the program output.
Any input that resulted in the assertion failing was considered to have discovered the information leak.

We had expected that DFSan would be unable to detect the three implicit flows due to the fact that DataFlowSanitizer tracks only \emph{data flow} and not \emph{control flow}; this proved to be the case in all runs.
It was slightly less clear whether PostgreSQL would trigger an assertion failure due to the secret values being stored into the database before being fetched in certain queries that expose the information leak.
As the database might write the values to disk and retrieve from disk, we expect that any taint labels are lost in this process.
It is also possible that no query exposing the leak was generated, however it seems unlikely that this would be the case for all 20 runs given that \leakfuzzer{} succeeded in 11 of its 20 runs.
Whatever the cause, the results show that for the setup that we used, DFSan was unable to detect the information flow from secret input to public output.
Finally, we observe that the error in cpuset is discovered very quickly by DFSan; this is a simple program, and is explored by the fuzzer very quickly.

The NetworkManager benchmark is a better candidate for detection by DFSan than by MSan, however as both require all dependencies to be compiled with sanitizer flags it was not tested.
Attempts were made to resolve all dependencies however after significant time was spent attempting to do this all the way down the stack too many issues were encountered to consider going any further.
Note that the `banking', `password check' and `reviewers' benchmarks required building the C++ standard library from scratch too, though this was managed in reasonable time.

\noindent
{\bf C Bounded Model Checker (CBMC)}

Model checking harnesses were created for each of the benchmarks to allow the model checker to verify the presence of information leakage.
An issue was encountered in building the `banking', `password check' and `reviewers' benchmarks, as these are written in C++.
This is due to a known issue that CBMC parser ``does not handle more modern and template-heavy C++, this means it often runs into problems with parts of the standard library'' \cite{cbmcCppIssue}. 
As a result, the model checker was evaluated on the 7 benchmarks written in C.

\vspace*{-0.2cm}
\begin{table}[H]
    \caption{Results for CBMC on the SIFF benchmark suite.}
    \vspace*{-0.3cm}
    \begin{small}
		\begin{tabular}{llp{0.8cm}p{1.0cm}l} \toprule
			\multirow{2}{*}{SUT} & \multirowcell{2}{\% runs leak\\detected} & \multicolumn{2}{l}{Leak detect time (s)} & \multirowcell{2}{Exit Cause} \\
			&                                                    & Mean      & Std. Dev   &   \\
			\midrule
			appletalk & 100 & 0.29 & 0.10 & Success \\
			cpuset    & 100 & 1.21 & 0.38 & Success \\
			
			NetworkManager   & 0            & -             & -                &    Segfault   \\
			OpenSSL-1.0.1f      & 0            & -             & -                &    Timeout   \\
			PostgreSQL           & 0            & -             & -                &     Timeout   \\
			RDS                         & 0            & -             & -                &    OOM \\
			sr9700\_rx\_fixup & 0            & -             & -                &    Timeout   \\ \bottomrule
		\end{tabular}
    \end{small}
\end{table}
\vspace*{-0.2cm}

The two smallest benchmarks, appletalk and cpuset, both finished quickly and CBMC managed to find a counterexample falsifying the assertion.
In one of these cases, it was able to detect the leakage faster than \leakfuzzer{} on average.
\leakfuzzer{} had a mean discovery time for appletalk of 175.23s, compared to CBMC's 0.29s.

CBMC struggled with larger programs as again a 24-hour time limit was imposed, leading to three subjects consistently exiting before the model checker was finished (marked in the table as `timeout').
In the case of RDS, the 128GB of memory available on the evaluation hardware was exhausted causing an out of memory (OOM) early exit.
Finally, when attempting the verify NetworkManager, CBMC would exit due to a segmentation fault within itself. 

The formal verification approach was faster in the instances where it managed to resolve, however as has been found in prior works it fails to scale well enough to work on the larger programs.
\\

\vspace*{-0.2cm}
\noindent
{\bf Memory Usage Comparison}

\begin{table}[H]
     \vspace*{-0.2cm}
    \caption{Table of results for CBMC}
     \begin{small}
		\begin{tabular}{lp{1.5cm}p{1.5cm}p{1.5cm}} \toprule
			\multirow{2}{*}{SUT} & \multicolumn{3}{l}{Mean Memory Usage (MiB)} \\
			                                    & \leakfuzzer{}  & AFL++  & CBMC   \\
			\midrule
            appletalk                     & 2,671.68        & 28.73    & 19.13    \\
            banking                       & 22,699.78     & 28.80    & -            \\
            cpuset                         & 1,926.38         & 28.74    & 12.22    \\
            NetworkManager       &  1,622.63        & 28.27    &  1,633.14    \\
            OpenSSL-1.0.1f          & 3,196.45         & 37.75    & 66,685.88 \\
            Password Check        & 603.52            & 29.03    & -  \\
            PostgreSQL                & 710.54            & 4.27       & 7,734.69   \\
            RDS                             & 7,433.65        & 32.30     & 127,203.04  \\
            Reviewers                   & 423.30            & 36.76    & - \\
            sr9700\_rx\_fixup             & 5,949.39       & 205.25   & 22,477.16 \\ \bottomrule
		\end{tabular}
    \end{small}
    \vspace*{-0.4cm}
\end{table}

One of the concerns with storing information about every tested input in \leakfuzzer{} as described in section \ref{sec:hashmap} is the potential memory usage.
As can be seen, over the 24 hour runs, memory usage of \leakfuzzer{} is certainly much higher than AFL++ with the sanitizers, but typically lower than CBMC in the long running test cases.
None of the 200 \leakfuzzer{} runs ran out of memory on the evaluation machine equipped with 128GB RAM.

\vspace*{0.1cm}
\noindent
\fbox{
	\begin{minipage}{0.99\columnwidth}
		\leakfuzzer{} was able to detect the information leak in all 10 of the programs in some proportion of the 24-hour runs.
		The combination of AFL++ with the sanitizers was able to detect leakage in three subjects, and CBMC was able to detect leakage in just two subjects.
		Of the 10 programs, information leakage was detected by all methods in just one: cpuset.
		Additionally, \leakfuzzer{} has been shown to use more memory than AFL++ as expected, however generally less than CBMC, and there were no issues with memory exhaustion in our tests.
	\end{minipage}
}

%% file: conclusion.tex
\section{Conclusions and Future Work}

We have presented \leakfuzzer{}, a fuzzer-based hypertesting approach to detecting information leakage, and proven its ability to do so on a varied range of benchmarks including seven information leakage related CVEs.
It has outperformed a combination of state of the art existing error detecting techniques, and demonstrated its ability to scale to real-world systems of considerable size.

In the future, we hope to extend \leakfuzzer{} to handle programs written in other programming languages, where sanitizers are not available.
Additionally we plan to quantify the information leakage to allow for application to a broader range of programs, such as password checkers, that need some level of information leakage in order to function usefully.

\section{Data Availability}

A replication package including results of the evaluation in this paper is freely available at \url{https://figshare.com/s/fa143b65420ab7ab1e3c}. 